\documentclass[preprint,prd,amsmath,amssymb,nofootinbib,tightenlines,floatfix,letterpaper]{revtex4}
\usepackage[dvips]{graphics}
\usepackage{epsfig}
\usepackage{latexsym}
\usepackage{slashed}
\newcommand{\be}{\begin{equation}}
\newcommand{\ee}{\end{equation}}
\newcommand{\bea}{\begin{eqnarray}}
\newcommand{\eea}{\end{eqnarray}}

\begin{document}
\preprint{NUHEP-TH/11-01}

\title{Flavored Dark Matter in Direct Detection Experiments and at LHC}

\author{Jennifer Kile\footnote{
	Electronic address: jenkile@northwestern.edu}
}
\affiliation{Department of Physics and Astronomy, Northwestern University, Evanston, IL 60208 USA}

\author{Amarjit Soni\footnote{
	Electronic address: soni@bnl.gov}}
\affiliation{Physics Department, Brookhaven National Laboratory, Upton, NY 11973 USA}

\begin{abstract}
We consider the possibility that dark matter can communicate with the Standard Model fields via flavor interactions.  We take the dark matter to belong to a ``dark sector'' which contains at least two types, or ``flavors'', of particles and then hypothesize that the Standard Model fields and dark matter share a common interaction which depends on flavor.  As, generically, interaction eigenstates and mass eigenstates need not coincide, we consider both flavor-changing and flavor-conserving interactions.  These interactions are then constrained by meson decays, kaon mixing, and current collider bounds, and we examine their relevance for direct detection and LHC.  
\end{abstract}

\maketitle

\section{Introduction}
\label{sec:intro}

Although the case for dark matter (DM) is now quite convincing (for reviews summarizing the evidence for DM, see, for example, \cite{Freese:2008cz,Bertone:2004pz}), very little is yet known about its identity.  The mass or masses of dark matter particles remains essentially unconstrained.  Aside from limits from direct detection experiments \cite{Aalseth:2010vx,Ahmed:2009zw,Aprile:2011hi,Ahmed:2010wy,ed:2011cy} and from observations of the bullet cluster \cite{Randall:2007ph}, we have essentially no data on DM interactions with itself or with the Standard Model (SM).  Meanwhile, the list of possible DM candidates is extensive and includes the lightest supersymmetric particle, axions, sterile neutrinos, and Kaluza-Klein DM; an extensive review of traditional DM candidates is given in \cite{Feng:2010gw}.

Perhaps most importantly, we do not even know how many species of dark matter there are, and, even if DM is comprised of just one species of particle, we do not know whether or not that particle is just the lightest in some ``dark sector''.  In recent years, we have seen an increasing number of models with more complex DM scenarios, such as two-component DM \cite{Fairbairn:2008fb}, multicomponent DM \cite{Zurek:2008qg}, exciting DM \cite{Finkbeiner:2007kk}, inelastic DM \cite{TuckerSmith:2001hy}, and related models \cite{ArkaniHamed:2008qn}.  Given this state of affairs, it makes sense to consider models with multiple DM components or novel DM-SM interactions. 

At the same time, the SM leaves unanswered many questions regarding the quark and lepton flavors.  Although in the last few years the precision with which the elements of the Cabibbo-Kobayashi-Maskawa (CKM) matrix are known has improved significantly \cite{Amsler:2008zzb}, there remains a need for better understanding of the triplication of the fermion generations in the SM, or how the three fermion generations could be related through some physics beyond the SM (BSM).  (For reviews of quark flavor physics, see \cite{Grossman:2010gw,Perez:2009xj,Nir:2007xn}.)  

Here, we consider the possibility that BSM physics which controls interactions between DM and the SM fields and the new physics which explains the flavor structure of the SM may be related.  We assume that DM belongs to a dark sector which contains at least two ``flavors'' of particles, $f$ and $f'$, both of which we take to be fermionic.\footnote{For a model of DM which has implications for flavor physics, but where the DM itself does not carry flavor, see \cite{Kilic:2010fs}.}  We take $m_f<m_{f'}$ and assume that DM is comprised of either $f$ or some mixture of $f$ and $f'$.  We will consider a wide range of DM masses, $m_f\lesssim \mbox{TeV}$. 

Given the lack of information on dark matter interactions, and the wide range of flavor models available, we find it appropriate to consider SM-DM interactions from a phenomenological point of view, instead of trying to incorporate DM into a specific model of flavor.  For this work, we concentrate on flavor interactions which involve the SM quark fields; as an illustration here, we will concentrate mainly on $d$ and $s$ quarks.  We consider two example interactions:  1) the case where both the SM quarks and the DM fields interact through purely vector couplings, and 2) where both interact through purely right-handed interactions.  Generically, one would expect both flavor-conserving and flavor-changing interactions; we include both possibilities here.

The layout of this paper is as follows.  First, in Sec. \ref{sec:not}, we introduce the idea of flavored DM and specify our notation and assumptions.  In Sec. \ref{sec:flav}, we review the current constraints on flavor interactions involving DM and $d$ and $s$ quarks from low-energy measurements and collider experiments and explore implications of the relic density on the interactions of flavored DM with the SM.  In Sec. \ref{sec:direct}, we consider limits from direct-detection experiments, taking into account the possibility that the dark sector may contain more than one long-lived component.  Next, we present two toy models of flavor gauge interactions in Sec. \ref{sec:toy}.  The relevance of TeV-scale flavor and DM interactions to LHC is explored in Sec. \ref{sec:lhc}.  Finally, in Sec. \ref{sec:conc}, we conclude.

\section{Notation}
\label{sec:not}
In this section, we will specify the interactions which we will consider among $s$ and $d$ quarks and our dark sector particles $f$ and $f'$.  For low-energy observables, we will primarily be interested in effective operators of the form
\begin{equation}
\label{eq:opstruct2}
\frac{C_{mnab}^g}{\Lambda^2}{\cal O}_{mnab}^g = \frac{C_{mnab}^g}{\Lambda^2} (\bar{f}_{m} \Gamma^{\mu} f_{n})(\bar{q}_{a} \Gamma_{\mu} q_{b})  
\end{equation}
which give interactions between the SM quarks and the dark sector.  We will also occasionally consider the four-quark operators
\begin{equation}
\label{eq:opstruct1}
\frac{C_{abcd}^{g}}{\Lambda^2}{\cal O}_{abcd}^{g} = \frac{C_{abcd}^{g}}{\Lambda^2} (\bar{q}_{a} \Gamma^{\mu} q_{b})(\bar{q}_{c} \Gamma_{\mu} q_{d}).  
\end{equation}
In these operators, the indices $m$ and $n$ indicate whether we are talking about $f$ or $f'$, while $a$, $b$, $c$, and $d$ on the quark fields $q$ indicate whether the quark flavor is $s$ or $d$.  We will assume that these flavor interactions are mediated by heavy gauge bosons (which we will denote generically as $Z'$), and, thus, we will confine our discussion to the example cases of purely vector ($g=V$) interactions with $\Gamma^{\mu}=\gamma^{\mu}$ and purely right-handed ($g=R$) interactions with $\Gamma^{\mu}=\gamma^{\mu}(1+\gamma^5)/2$.  

Each of these operators is multiplied by a coefficient $C_{mnab}^{g}/\Lambda^2$ (or $C_{abcd}^{g}/\Lambda^2$) where $\Lambda$ is taken to be some high new physics scale.  As we assume that the SM and DM share a common flavor interaction, we expect these operators to have similar scales.  However, the operator coefficients may also contain small mixing angles; in the SM weak interaction, these small mixing angles cause the effective scales between different four-quark operators to differ by more than $2$ orders of magnitude.  Here, we will keep all of our results in terms of $C_{mnab}^{g}/\Lambda^2$ and $C_{abcd}^{g}/\Lambda^2$.

For some parts of this analysis, we will have to also consider possible interactions involving other quark fields.  In the case of vector interactions, we must include both right-handed quarks (which are singlets under the SM $SU(2)$ weak interaction), as well as the left-handed $SU(2)$ doublets; this necessarily requires that we also consider up-type quarks, weighted by the appropriate angles of the CKM matrix.  Additionally, the interactions we have specified above, if taken in isolation or without careful arrangement of quantum numbers, lead to anomalies in triangle diagrams involving the $Z'$s and the SM gauge bosons; however, as the choice of quantum numbers for the SM and dark sector fields needed to cancel these anomalies is not unique, we will not consider these additional interactions throughout most of this paper.  

Finally, among the operators included in Eq. (\ref{eq:opstruct1}), there exist some (${\cal O}_{sdsd}^g$ and their Hermitian conjugates) which change strangeness by two units, and, thus, can contribute to $K^0-\bar{K}^0$ mixing.  Constraints on the effective new physics scale for these operators are $O(10^3\mbox{ TeV})$ \cite{Nir:2007xn}.  As we are interested in effects which may be observed at LHC or DM direct detection experiments, our analysis will only be applicable to flavor models which do not have tree-level contributions to $K^0-\bar{K^0}$ mixing.  In Sec. \ref{sec:toy}, we will present two toy models, one with right-handed couplings and one with vector couplings, which are anomaly-free and which do not contribute to $K^0-\bar{K^0}$ mixing at tree level.

\section{Constraints on Flavored Dark Matter }
\label{sec:flav}
In this section, we will review some of the relevant measurements which constrain BSM flavor interactions involving $d$ and $s$ quarks with each other and with the dark sector particles $f$ and $f'$.  We begin with constraints obtainable from low-energy observables.  We first consider the case where the $f$ (and possibly also the $f'$) is very light, $m_f\lesssim 180$ MeV.  In this case, we would expect to have the decay $K^+ \rightarrow \pi^+ f\bar{f}$ (and possibly decays to final states containing $f'$ or $\bar{f}'$ as well).  Thus, considering the branching fraction of $K^+$ to a $\pi$ plus neutrinos \cite{Amsler:2008zzb}
\begin{equation}
\mbox{Br}(K^+ \rightarrow \pi^+ \nu\bar{\nu}) = 1.7 \pm 1.1 \times 10^{-10} 
\end{equation}
and taking the $2 \sigma$ bound on this measurement as a limit on the branching fraction to $f\bar{f}$, and taking the ratio of this branching fraction to that of the SM process $K^+\rightarrow e^+ \nu \pi^0$, we obtain, for purely right-handed interactions, 
\begin{equation}
\frac{|C_{mnsd}^R|}{\Lambda^2} < \frac{1}{(47\mbox{ TeV})^2}
\end{equation}
for $m_f$ (and possibly also $m_{f'}$) $<<180\mbox{ MeV}$.  We can also consider the case where $f$ is very light, but $m_{f'}$ is somewhat heavier, by comparing to the SM process $K^+\rightarrow \mu^+ \nu \pi^0$, obtaining 
\begin{equation}
\frac{|C_{ff'sd}^R|}{\Lambda^2} \lesssim \frac{1}{(42\mbox{ TeV})^2}
\end{equation}
for $m_{f'}\approx m_{\mu}$.  (Here, we assume that the limit on $K^+ \rightarrow \pi^+ f\bar{f}$ is not substantially degraded when one of the final-state dark sector particles acquires a mass of $O(100\mbox{ MeV})$.)

We see from these bounds that the case of very light $f$ is very strongly constrained.  (Although in both cases here we assumed that $f$ was very light, $m_f<<180$ MeV, we can infer from the strength of these constraints that we would still obtain significant bounds on the new physics scale for all cases in which $K^+$ decay was not strongly phase-space suppressed.)  The limits derived here are only applicable to the operators ${\cal O}_{mnsd}^R$ as shown above (and, unless $m=n=f$, only to the case where both $f$ and $f'$ are light).  As interaction and mass eigenstates need not coincide, we may be tempted to interpret these results as a tentative order-of-magnitude estimate of expected bounds on the new physics scale for the operators ${\cal O}_{mndd}^R$ and ${\cal O}_{mnss}^R$ as well; however, small mixing angles ($\lesssim 0.1$) can easily invalidate this interpretation.

Our results are similar for the case of vector interactions, 
\begin{equation}
\frac{|C_{mnsd}^V|}{\Lambda^2} \lesssim \frac{1}{O(80\mbox{ TeV})^2}
\end{equation}
for the case where both final-state dark sector particles are light, and 
\begin{equation}
\frac{|C_{mnsd}^V|}{\Lambda^2} \lesssim \frac{1}{O(70\mbox{ TeV})^2}
\end{equation}
for the case where one is light and the other has a mass of $O(100\mbox{ MeV})$.  In this case, however, because vector interactions with quarks necessarily involve the left-handed quark doublets, and because the interaction eigenstates in the left-handed doublets differ from the mass eigenstates by a rotation via the Cabibbo angle $\theta_C$, the operators ${\cal O}_{mnab}^V$ necessarily involve some significant quark mixing.  Since we have no good reason to believe that the interaction eigenstates would be aligned with the {\it down-type} quarks, it therefore may be more compelling in this case to interpret the bounds on $\frac{|C_{mnsd}^V|}{\Lambda^2}$ as a tentative bound on the $a=b=d,s$ cases as well.  

We note that one can also obtain limits on the right-handed operators for the case $a=b=d$ from limits on supernova cooling, assuming that $m_f\lesssim \mbox{few}\times 10 \mbox{ MeV}$.  We can apply the limit on the branching fraction $\mbox{Br}(\pi^0\rightarrow \nu\bar{\nu})<3.2\times 10^{-13}$ \cite{Natale:1990yx,Amsler:2008zzb} to the case of a light $f$.  For $m_f\approx \mbox{few}\times 10\mbox{ MeV}$, we obtain limits on the new physics scale of order $\sim 1$ TeV.  However, as the decay $\pi^0\rightarrow f\bar{f}$ is helicity-suppressed, this limit quickly degrades with decreasing $m_f$.  Similar statements apply in the case that one or both final-state particles is an $f'$.

\begin{figure}[h]
\includegraphics[width=.5\textwidth, angle=0]{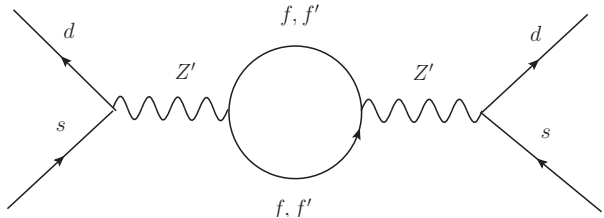}
\caption{$K^0-\bar{K}^0$ mixing contribution from internal dark fermion loop.}
\label{fig:kaonmix}
\end{figure}

We now move on the the case of heavier $f$.  We will first consider constraints which can be obtained from $K^0-\bar{K}^0$ mixing.  As we limit ourselves to models which have no tree-level contribution to $K^0-\bar{K}^0$ mixing, we will only concern ourselves with contributions from the diagram shown in Fig. \ref{fig:kaonmix}.  (In the case of vector interactions, there are also loop diagrams containing up-type quarks which can contribute to $K^0-\bar{K}^0$ mixing.  We will briefly discuss these contributions when we consider toy models in Sec. \ref{sec:toy}.)  For simplicity, we will only consider the case where $m_f, m_{f'}>>m_K/2$.

First, we will consider operators which change flavor on both the SM and dark sector fields.  For a right-handed interaction, we have
\begin{equation}
\frac{C_{f'fds}^R}{\Lambda^2}{\cal O}_{f'fds}^R = \frac{C_{f'fds}^R}{\Lambda^2} (\bar{f}_{R}' \gamma^{\mu} f_{R})(\bar{d}_{R} \gamma_{\mu} s_{R})  
\end{equation}
while for a vector interaction, we have
\begin{equation}
\frac{C_{f'fds}^V}{\Lambda^2}{\cal O}_{f'fds}^V = \frac{C_{f'fds}^V}{\Lambda^2} (\bar{f}' \gamma^{\mu} f)(\bar{d} \gamma_{\mu} s).  
\end{equation}
First, we note that these interactions, taken in isolation, do not contribute to the diagram in Fig. \ref{fig:kaonmix}; the $f$ created in one of the vertices in the loop would have to transform into an $f'$ before it is destroyed at the other vertex.  However, we note that, in general, we would not expect the interaction eigenstates in the dark sector to necessarily coincide with the mass eigenstates $f$ and $f'$.  We can introduce interaction eigenstates $f_1$ and $f_2$, related to $f$ and $f'$ by
\begin{equation}
\left( \begin{array}{c}
f\\
f'\end{array}\right)
=\left( \begin{array}{rr}
\alpha & \beta\\
-\beta^* & \alpha^*\end{array}\right)
\left( \begin{array}{c}
f_{1}\\
f_{2}\end{array}\right)
\label{eq:mix}
\end{equation}
where $|\alpha|^2+|\beta|^2=1$ and, in the case of the right-handed interaction, the fields in Eq. (\ref{eq:mix}) are taken to be the right-handed components only.  (Here, we assume that $f_{1,2}$  do not mix with any other states.)  We can now place limits on combinations of the coefficients $C^g_{f_2f_1ds}/\Lambda^2$, the parameters $\alpha$ and $\beta$ which control the mixing, and the mass splitting, $\delta=m_{f'}-m_f$.  In the case of right-handed interactions, the diagram in Fig. \ref{fig:kaonmix} is finite; we obtain, for the mass difference between $K_L$ and $K_S$,
\begin{equation}
\Delta_{m_K} = \left|A f_K^2 m_K \alpha^{*2} \beta^2 \left(\frac{C^{R}_{f_2f_1ds}}{\Lambda^2}\right)^2\frac{1}{(4\pi)^2} \delta^2\right|
\end{equation}
where $f_K\approx 160$ MeV is the kaon decay constant and $A$ is a constant which depends on the relative size of $m_f$ and $m_{f'}$; for $m_f\approx m_{f'}$, $A\approx 8/9$, while, for $m_f<< m_{f'}$, $A\approx 2/3$.  We can then compare this to the experimental value $\Delta_{m_K} = 3.48\times10^{-15}$ GeV \cite{Amsler:2008zzb}; for $\Lambda= 1\mbox{ TeV}$, we obtain upper bounds on $|C^R_{f_2f_1ds}\alpha^*\beta \delta|$ of approximately $7-8$ GeV. 

We get a similar order-of-magnitude expression for the case of vector interactions,
\begin{equation}
\Delta_{m_K} \approx \left|\frac{8}{3} f_K^2 \frac{m_K^3}{(m_s + m_d)^2}\alpha^{*2} \beta^2 \left(\frac{C^{V}_{f_2f_1ds}}{\Lambda^2}\right)^2\frac{1}{(4\pi)^2} \delta^2 \ln\left(\frac{\Lambda^2}{m_{f'}^2}\right)\right|
\end{equation}
where the logarithmic behavior comes from the running of $C^V_{dsds}$ between the new physics scale and $m_{f'}$ induced by the diagram in Fig. \ref{fig:kaonmix}.  For $\Lambda= 1 \mbox{ TeV}$, this gives an upper bound on $|C^V_{f_2f_1ds}\alpha^*\beta \delta|$ of $\approx 1$ GeV.  Here, we have taken the logarithmic factor in the vector case to be of order unity, but we note that the upper bound on $|C^V_{f_2f_1ds}\alpha^*\beta \delta|$ can be strengthened considerably if $f$ and $f'$ are very light.  Thus, we see that fairly small mass splittings are phenomenologically interesting in these scenarios.   

We also briefly consider the operators which do not change dark sector flavor.  For both the right-handed and vector interactions, the diagram in Fig. \ref{fig:kaonmix} will be quadratically divergent.  If we regulate this divergence using dimensional regularization, we find that the contribution to $\Delta_{m_K}$ from the operator for the right-handed case,
\begin{equation}
\frac{C_{ffds}^R}{\Lambda^2}{\cal O}_{ffds}^R = \frac{C_{ffds}^R}{\Lambda^2} (\bar{f}_{R} \gamma^{\mu} f_{R})(\bar{d}_{R} \gamma_{\mu} s_{R})  
\end{equation}
will contain factors of $m_f$ instead of $\delta$ and, thus, this operator will be more strongly constrained than the operators which change dark sector flavor as long as $m_f$ is substantially larger than $\delta$.  The analogous vector operator, on the other hand, will contain factors of $m_K$ instead of $\delta$ or $m_f$; the scale for this operator will be constrained to be at least $O(\mbox{TeV})$. 

We also mention that the limits derived from $\Delta_{m_K}$ above depend only on the magnitude of $C^g_{f_2f_1ds}\alpha^*\beta$, not its phase.  However, the CP-violation parameter $\epsilon_K$ is sensitive to a complex phase in the kaon mixing matrix, which depends on $C^g_{f_2f_1ds}\alpha^*\beta$.  Depending on the choice of these phases, the upper bound on $|C^g_{f_2f_1ds}\alpha^*\beta\delta|$ could be strengthened by more than an order of magnitude for a given value of $\Lambda$.

We briefly mention relevant collider bounds.  CDF \cite{Aaltonen:2008dn} has directly looked for heavy neutral gauge bosons, $Z'$, decaying to jets.  They have excluded the mass range of $320-740$ GeV, assuming SM couplings.  (For limits on lighter $Z'$ gauge bosons, see results from UA2 \cite{Alitti:1990kw,Alitti:1993pn}.)  However, these limits can be easily evaded by making the couplings between the quarks and the $Z'$ slightly smaller than those in the SM.  Additionally, indirect limits on specific flavor models from fits to collider data were given in \cite{Burdman:1999us}.

Finally, we consider constraints obtained from the observed DM density of $\Omega_{dark}=0.228\pm 0.013$ \cite{Hinshaw:2008kr}.  If DM consists of only one species, obtaining the correct relic density requires a velocity-averaged annihilation cross-section at freezeout $<\sigma v_r>$ of $\approx 3 \times 10^{-26} \mbox{cm}^3/\mbox{s}$, where $v_r$ is the relative speed of DM particles, with a mild dependence on $m_f$.  As long as $m_f\gtrsim O(\mbox{GeV})$, each of our purely right-handed operators will contribute a term
\begin{equation}
<\sigma v_{r}>_{mnab}^R \approx \frac{|C_{mnab}^R|^2}{\Lambda^4} \frac{3 m_f^2}{8\pi}
\end{equation}
to the annihilation cross-section $f_m \bar{f}_n \rightarrow q_a \bar{q}_b$.  (Here, we have neglected velocity-dependent terms.)  This gives a value for the sum of the squares of the operator coefficients, 
\begin{equation}
\sum \frac{|C_{mnab}^R|^2}{\Lambda^4} \approx \frac{1}{(2.6 \mbox{ TeV})^4}\left(\frac{\mbox{TeV}}{m_f}\right)^2.
\label{eq:relr}
\end{equation}
We see that the new physics scale decreases with decreasing $m_f$.  We note, however, that, if we allow $f$ to comprise only a fraction of DM (and thus have a smaller relic density), we can allow larger values for $<\sigma v_r>$, and thus lower values for the new physics scale.  Thus, if we have only one operator with a nonzero coefficient, say, ${\cal O}^R_{ffdd}$, then $2.5$ TeV should be regarded as an approximate {\it upper} bound on the new physics scale.  (Note that all of our other constraints place {\it lower} bounds on the scale of new physics.)  Of course, this upper bound on the scale of new physics can be raised by having additional terms with nonzero coefficients in the sum shown in Eq. (\ref{eq:relr}), so this should not be taken as a rigorous upper bound on the NP scale.

We get a similar term
\begin{equation}
<\sigma v_r>_{mnab}^V = \frac{|C_{mnab}^V|^2}{\Lambda^4} \frac{6 m_f^2}{\pi}
\end{equation}
for the operators with purely vector interactions.  Here, we have included a factor of $2$ in the cross-section, as gauge invariance demands that we include both the upper and lower components of the left-handed quark doublets.  This gives
\begin{equation}
\sum \frac{|C_{mnab}^V|^2}{\Lambda^4} \approx \frac{1}{(5.2 \mbox{ TeV})^4}\left(\frac{\mbox{TeV}}{m_f}\right)^2.
\label{eq:rell}
\end{equation}

We note that our relic density calculations assume the simplest scenario for DM self-annihilation $f\bar{f}\rightarrow q\bar{q}$.  If $f$ and $f'$ have masses which differ by only a few percent or less, the more complicated coannihilation calculation \cite{Griest:1990kh} is relevant; coannihilations can significantly reduce the relic density of $f$, and thus loosen the limits in (\ref{eq:relr}) and (\ref{eq:rell}), if the annihilation cross-section for $f'\bar{f'}$ is substantially larger than that for $f\bar{f}$.  Additionally, this calculation also neglects the possibility of other annihilation channels for $f\bar{f}$, which can substantially raise the new physics scale.

\section{Signatures in Dark Matter Direct Detection Experiments}
\label{sec:direct}
We now discuss the prospects for direct detection of flavored DM.  We will distinguish between two experimentally distinct cases, depending on the mass splitting $\delta$.  In the first case, $\delta \gtrsim \mbox{ few}\times 100 \mbox{ keV}-1 \mbox{ MeV}$.  (The range of masses given here reflects the difference between the right-handed and vector operators.)  In this case, $f$ is the only long-lived particle in the dark sector, and it interacts elastically with nucleons, $fN\rightarrow fN$.  In the second case, as we will show below,  it is possible that the heavier state $f'$ can also be long-lived and form a sizable component of DM, thus allowing exothermic down-scattering of the form $f' N \rightarrow f N$.  (Additionally, for $\delta \lesssim \mbox{ few}\times 100$ keV, up-scattering of the form $fN\rightarrow f'N$ becomes possible, thus making flavored DM a possible example of inelastic Dark Matter \cite{TuckerSmith:2001hy}, introduced to possibly explain the apparent conflict between DAMA \cite{Bernabei:2010zza} and other experiments.  Although this scenario is significantly constrained \cite{Ahmed:2009zw,Angle:2009xb,Akimov:2010vk}, it remains a viable possibility for explaining the DAMA results if $\delta\sim O(200\mbox{ keV})$ \cite{Chang:2010pr}.\footnote{We note, however, that arranging for up-scattering to occur while suppressing the elastic interaction $fN\rightarrow fN$ presumably requires significant fine-tuning of mixing angles; \cite{Chang:2010pr} utilize inelastic scattering cross-sections which are $10$ orders of magnitude larger than the elastic scattering cross-sections ruled out by \cite{Aprile:2011hi} to address the DAMA results.  This does not preclude, however, the existence of inelastic scattering with much smaller cross-sections, which could require less fine-tuning of mixing angles.})  

First, we will discuss the case where $\delta$ is sufficiently large that $f'$ is not long-lived, and DM can interact in direct detection experiments only through the elastic reaction $f N \rightarrow f N$.  In this case, we can directly apply the constraints from existing DM experiments; here, we will only consider spin-independent contributions to the cross-section, as they are significantly more strongly constrained than spin-dependent contributions.  As direct detection experiments search for interactions between DM and nucleons, we are interested in operators which contain $d$ quarks (and $u$ quarks, in the case of vector interactions).  For the purely right-handed interactions, the measured spin-independent DM-nucleon cross-section takes the form
\begin{equation}
\sigma_{SI}=\frac{|C^R_{ffdd}|^2}{\Lambda^4}\frac{1}{16\pi}M^2_{red}\frac{(Z+2(A-Z))^2}{A^2},
\end{equation}
where $M_{red}$ is the reduced mass of the DM-nucleon system and $A$ and $Z$ are the atomic mass number and atomic number of the target nucleus, respectively.  The dependence on $A$ and $Z$ takes into account the fact that limits from DM direct-detection experiments assume that the cross-sections for DM scattering on protons and neutrons are equal; $f$ interacts only with $d$ quarks, and not $u$ quarks; thus, the cross-section for a neutron is four times that of a proton.  For the purely vector interactions we have
\begin{equation}
\sigma_{SI}=\frac{|C^V_{ffdd}|^2}{\Lambda^4}\frac{9}{\pi}M^2_{red}
\end{equation}
where the factor of $9$ includes the contributions from both $u$ and $d$ quarks.  (As $f$ interacts identically with $u$ and $d$ quarks, the cross-sections here are the same for protons and neutrons; we ignore corrections due to the Cabibbo angle.)  

We can compare these expressions to the cross-section limits from direct detection experiments; here we will assume that all of DM is comprised of $f$.  We consider three specific values for $m_f$. For the CoGeNT signal region, $m_f\approx 10$ GeV \cite{Aalseth:2010vx}, we take $\sigma_{SI}\approx 5\times 10^{-41}\mbox{ cm}^2$ (although this is in conflict with \cite{Aprile:2011hi,Ahmed:2010wy}).  For $m_f\approx1$ TeV and the range of DM masses where these limits are strongest, $m_f\approx {\cal O}(10\mbox{'s of GeV})$, we use the results from XENON100 \cite{Aprile:2011hi}, who report a spin-independent cross-section upper bound for these mass ranges of $\sigma_{SI}\lesssim 8\times 10^{-44}\mbox{ cm}^2$ and $\sigma_{SI}\lesssim 7\times 10^{-45}\mbox{ cm}^2$, respectively.
Our results are shown in Table \ref{table:dir}.

\begin{table}[h]
\begin{tabular}{| c | c | c |}
\hline
$m_f$ & $(|C_{ffdd}^R|/\Lambda^2)^{-1/2}/\mbox{TeV}$ & $(|C_{ffdd}^V|/\Lambda^2)^{-1/2}/\mbox{TeV}$\\
\hline
$\sim10$ GeV & $\sim 0.7$ & $\sim 2$ \\
\hline
few$\times 10$ GeV & $\gtrsim 7$ & $\gtrsim 19$\\
\hline
$\sim1$ TeV & $\gtrsim 4$ & $\gtrsim 11$\\
\hline

\end{tabular}
\caption{Results for the new physics scale from DM direct detection experiments.  The first line refers to NP scale corresponding to the CoGeNT signal region \cite{Aalseth:2010vx}, while the last two are lower bounds on the NP scale obtained from the results of XENON100 \cite{Aprile:2011hi}.}
\label{table:dir}
\end{table}

It should be noted that these direct detection limits on the lower bound for the new physics scale disagree with the upper bounds from the relic density calculation unless $f$ interacts with more than just $d$ (and, in the case of the vector operators, one family of up-type) quarks.  Given the significant tension between the direct detection limits and the relic density bounds, it may be necessary to solve this problem by, for example, introducing annihilation channels of DM particles into leptons \cite{Kopp:2009et,Bell:2008vx}, introducing additional interactions and/or particles to the dark sector, or positing mechanisms which would give $f$ a non-thermal cross-section.  As these issues are not specific to flavored DM, we will not try to address them here; we will instead take the attitude that explaining the relic density will require the construction of a specific model.  (We also note that if we relax the assumption that $f$ comprises all DM, the direct detection constraints are also loosened.)

We now address the question of when the excited state $f'$ can be suitably long-lived to form a significant component of DM.  Here, we make the conservative assumption that the couplings of $\bar{f}f'$ to $d\bar{d}$ and $s\bar{s}$ are not significantly smaller than those to $d\bar{s}$ and $s\bar{d}$.  (If couplings which change SM flavor dominate, $f'$ decays will be suppressed by factors greater than those shown below for the flavor-diagonal case; this may happen, for example, in the toy models presented in Sec. \ref{sec:toy} if the DM interaction eigenstates are closely aligned with the mass eigenstates.)  For the case of right-handed interactions, $f'$ can decay at tree-level as $f'\rightarrow f \pi^0$ if $\delta>m_{\pi}$; tree-level decay for the vector coupling case can occur via $f'\rightarrow f + \mbox{ jets}$ for somewhat higher $\delta$.  In both of these cases, $f'$ will not be long-lived unless the scale of new physics is extremely high; thus, we consider the cases of substantially smaller $\delta$ and decays involving neutrinos and photons.  For the case of right-handed interactions, we consider $f'\rightarrow f \nu \bar{\nu}$ and decays containing one or two photons.  We will find that $f'$ decays are substantially more weakly constrained in the case of vector couplings; in this case, we will consider the decays $f'\rightarrow f \gamma\gamma\gamma$ and $f'\rightarrow f e^+ e^-$ as well.  We note that, while we are interested in the case where $f'$ lives long enough to comprise a significant fraction of DM today, there also exist constraints on DM particles with lifetimes substantially less than the age of the universe but greater than $O(1 \mbox{ second})$; see \cite{Kanzaki:2007pd} and references therein.

\begin{figure}[h]
\includegraphics[width=.25\textwidth, angle=0]{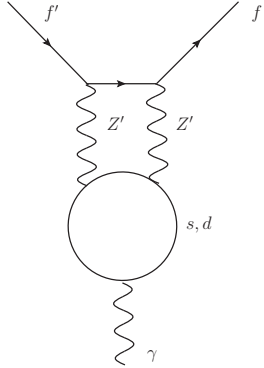}
\caption{Diagram contributing to $f'\rightarrow f \gamma$ in the case of right-handed couplings to the $Z'$.}
\label{fig:decay1}
\end{figure}

For the case of right-handed interactions, the strongest constraint we obtain is by considering the $f'\rightarrow f\gamma$ diagram shown in Fig. \ref{fig:decay1}.  This diagram is superficially logarithmically divergent; however, if we consider this diagram from an effective-operator point of view, gauge invariance requires that the decay occur via an effective transition magnetic moment operator, $\bar{f}\sigma^{\mu\nu}f'F_{\mu\nu}$.  As this operator contains an explicit factor of the photon momentum, the number of factors of loop momenta that we integrate over must be reduced by one; this renders the diagram finite.  We will assume that the integrals over the loop momenta can be reasonably estimated by cutting them off at our new physics scale $\Lambda$; for simplicity, we take the $C_{mnab}^R$ to be $O(1)$.

The diagram in Fig. \ref{fig:decay1} contains a sub-diagram which couples two $Z'$ bosons to a photon via a quark loop.   We take this sub-diagram to be similar to the fermion loop contribution to the SM effective $ZZ\gamma$ vertex \cite{Barroso:1984re}, which depends very weakly on the fermion mass in the scenario of large gauge boson momentum, as is relevant here.  However, this sub-diagram is anomalous (as will be discussed in more detail in Sec. \ref{sec:toy}).  Therefore, when we sum over all possible fermions in the loop, terms which are independent of fermion mass will cancel, and we must look at mass-dependent terms, which, roughly speaking, will contribute a factor of $m_q^2/\Lambda^2$ to the decay amplitude, where $m_q$ is the mass of the fermion in the loop.  We thus arrive at our estimate for the decay of $f'$ to $f\gamma$:
\begin{equation}
\Gamma\sim \frac{1}{(4\pi)^8}\alpha \left(\frac{m_q}{\Lambda} \right)^4 \frac{m_f^2\delta^3}{\Lambda^4}.
\end{equation}
We then utilize the results of \cite{Yuksel:2007dr}, who use the Milky Way $\gamma$ ray line search to constrain dark matter decays containing monoenergetic photons.  For the values of $\delta$ relevant here, they exclude such decays if $\Gamma\gtrsim 10^{-49} m_f$.  Here, we take $\Lambda=1$ TeV and $m_q=m_s$\footnote{We take the value of $m_S(1\mbox{ TeV})=47\pm^{14}_{13}$ from \cite{Xing:2007fb}; we note that the error on this number can probably be reduced using the recent results of \cite{Aoki:2010dy}, but we emphasize that here we are interested in an order-of-magnitude result for $\delta$.}; this latter choice is valid if the flavor charges of the SM are arranged to cancel the anomalies; if, instead, new (heavy) particles are added, substantially stronger constraints will be obtained.  For these values, we obtain
\begin{equation}
\delta\lesssim (1200\mbox{ keV}) \left(\frac{\mbox{GeV}}{m_f}\right)^{1/3}
\end{equation}
as an order-of-magnitude upper bound on the values of $\delta$ which will allow $f'$ to be long-lived.  Although this is an approximate limit, we would like to point out that considering $f'$ decaying via a virtual pion will give values for $\delta$ of the same order of magnitude.

\begin{figure}[h]
\includegraphics[width=.7\textwidth, angle=0]{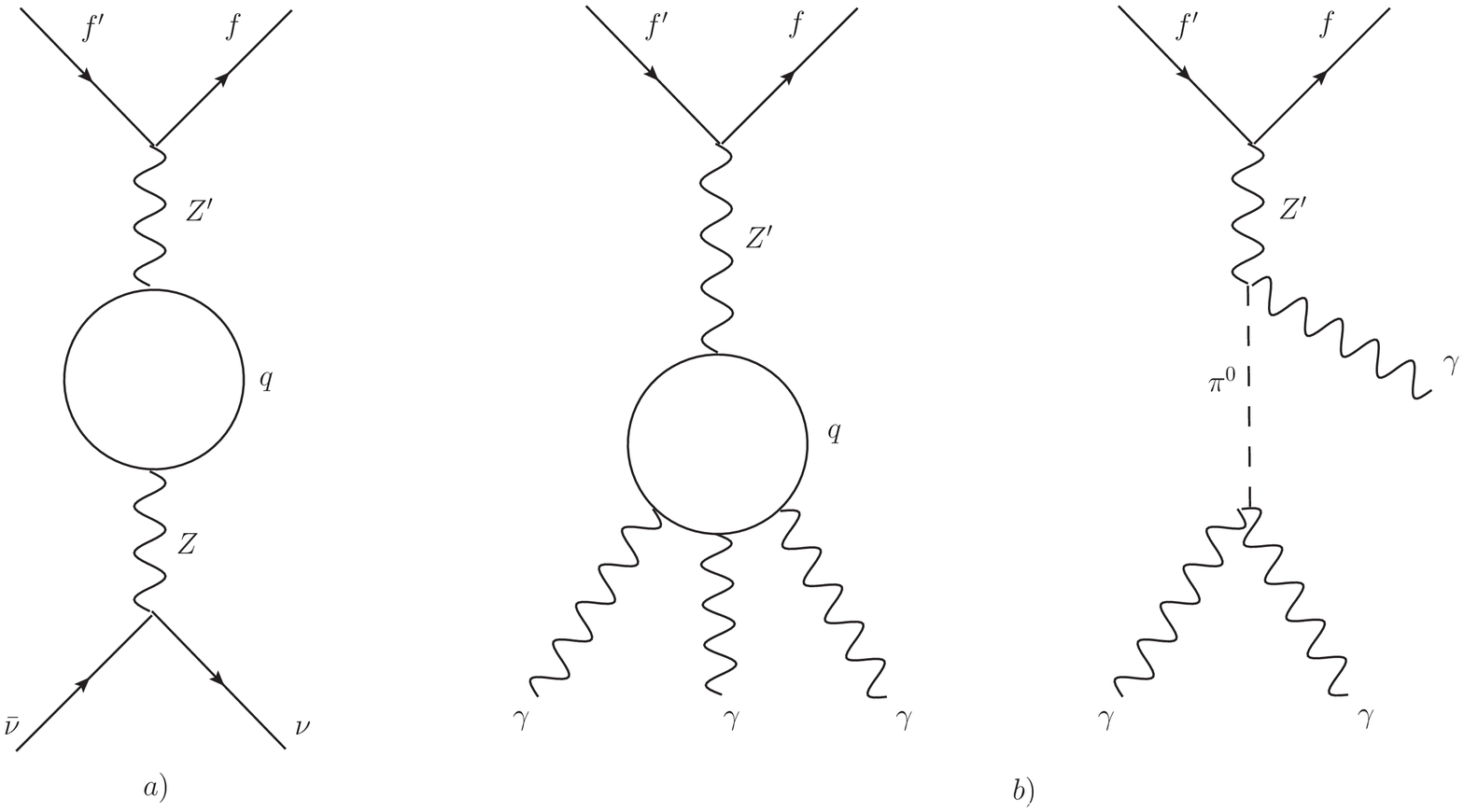}
\caption{Diagrams contributing to $f'$ decay in the case of vector couplings to the $Z'$.  a) $f'\rightarrow f \nu\bar{\nu}$ via $Z'-Z$ mixing.  b) Example diagrams contributing to $f'\rightarrow f \gamma\gamma\gamma$.}
\label{fig:decay2}
\end{figure}

For vector interactions, the $f'$ lifetime depends strongly on whether or not the decay channel $f'\rightarrow f e^+ e^-$ is kinematically allowed.  For $\delta<2 m_e$, only the decay channels $f'\rightarrow f \nu \bar{\nu}$ and $f'\rightarrow f + \mbox{photons}$ are allowed.  We considered many diagrams that could contribute to these decays.  The $Z'$ does not mix into an on-shell photon, so $f'\rightarrow f \gamma$ does not occur at one-loop order.  A quark loop connected to three vector bosons vanishes identically, which eliminates $f'\rightarrow f \gamma \gamma$ at one-loop order as well as $f'\rightarrow f \gamma$ via the two-loop diagram shown in Fig. \ref{fig:decay1}.  $f'\rightarrow f \gamma \gamma$ via a virtual $\pi^0$ does not occur for a purely vector interaction, and $f'\rightarrow f \gamma \gamma$ via a virtual $\rho$ is forbidden by charge-conjugation invariance.  The largest nonzero contributions to $f'$ decay via $Z'$s are $f'\rightarrow f \nu \bar{\nu}$ via $Z'-Z$ mixing and contributions to $f'\rightarrow f \gamma\gamma\gamma$ via a quark loop or virtual $\rho$ and/or $\pi^0$ mesons; example diagrams contributing to these processes are shown in Fig. \ref{fig:decay2}.  Both of these processes will be strongly suppressed; $f'\rightarrow f \nu \bar{\nu}$ will be suppressed by factors of both the $Z'$ and $Z$ masses, while the rate for $f'\rightarrow f \gamma\gamma\gamma$ is suppressed by many factors of the small photon momenta.  As in the case of right-handed couplings, if we do not add additional fermion fields, anomaly cancellation requires that coefficients of the operators with first and second generation quarks have equal magnitude and opposite sign, which renders the diagram in  Fig. \ref{fig:decay2}(a) finite.  If we take $\delta=1$ MeV and insist that $f'$ have a lifetime comparable to the age of the universe, this diagram gives limits on the new physics scale weaker than $O(\mbox{GeV})$.  (We note that it may be possible to slightly improve these limits using observations of dark matter halos \cite{Peter:2010au}.)  We also obtain order-of magnitude constraints on the new physics scale using diagrams for $f'\rightarrow f \gamma\gamma\gamma$ such as those in Fig. \ref{fig:decay2} (b) and using the limits on dark matter decays involving photons from \cite{Yuksel:2007dr}; for $\delta=1$ MeV and $m_f\sim100$ MeV, the limit on the new physics scale is no better than $O(10\mbox{ GeV})$; this limit weakens with growing $m_f$.

Although the limits from these diagrams are very weak, we note that it is possible to have the $f'$ decay in this scenario for $\delta<2 m_e$ and for interesting values of the new physics scale if we make some assumptions about the scalar sector of the model.  We will mention this briefly in Sec. \ref{sec:toy} when we discuss toy models. 

Finally, we mention for the case of vector interactions that if $\delta>2m_e$, $f'$ can decay much faster via a diagram similar to that in Fig. \ref{fig:decay2} (a) but with the SM $Z$ replaced by a photon and with the neutrinos replaced by an $e^+e^-$ pair.  

Thus, we conclude that for $\delta\lesssim\mbox{ few }\times 100\mbox{ keV}$ (for right-handed interactions) or $\delta\lesssim 1\mbox{ MeV}$ (for vector interactions), $f'$ can possibly be long-lived.  This opens up the possibility that $f'$ could be discovered via its distinctive down-scattering signatures at direct-detection experiments; for works where such signatures have been considered, see \cite{Batell:2009vb,Finkbeiner:2009mi,Alves:2010dd,Lang:2010cd,Graham:2010ca}.  Although these signatures would be useful only for very small $\delta$, if observed, they would provide strong evidence for multicomponent dark matter.

\section{Toy Models}
\label{sec:toy}
Here, we present two toy models which contain, respectively, the purely right-handed and purely vector couplings described above.  We show that they are anomaly-free and do not contribute significantly to $K^0-\bar{K}^0$ mixing.  In both cases, we assume that quarks from the first and second generations transform as a doublet under a gauged $SU(2)_F$ flavor symmetry, under which all other SM fields are singlets.  In the first case, only the right-handed down-type quarks transform under the $SU(2)_F$ flavor symmetry, while in the second case all quarks from the first and second generations transform under the symmetry.  In both scenarios, the $f$ and $f'$ also transform as a doublet under the same $SU(2)_F$ symmetry.  We now describe the particle content and couplings in each of these toy models.

\subsection{Right-handed couplings}
In this model, the right-handed down-type quarks form a doublet under $SU(2)_F$:
\begin{equation}
D_R=\left( \begin{array}{c}
d_{R1}\\
d_{R2}\end{array}\right)
\end{equation}
while all other SM fields are $SU(2)_F$ singlets.  Here, we perform a rotation such that $d_1$ and $d_2$ are aligned with the right-handed components of the mass eigenstates $d_R$ and $s_R$, respectively; we neglect possible mixing with $b_R$.  Similarly, we have
\begin{equation}
F_R=\left( \begin{array}{c}
f_{R1}\\
f_{R2}\end{array}\right)
\end{equation}
where we do not assume that $f_{R1}$ and $f_{R2}$ are necessarily closely aligned with the mass eigenstates $f_R$ and $f_R'$.  The three gauge bosons of $SU(2)_F$ are labeled $Z_i'$ for $i=1,2,3$; we assume that these gauge bosons obtain their mass through a scalar $SU(2)_F$ doublet $\varphi$ acquiring a vacuum expectation value which, for some $SU(2)$ rotation, can be written in the form
\begin{equation}
\varphi\rightarrow\left(\begin{array}{c} 0\\ \frac{v'}{\sqrt{2}}\end{array}\right).  
\end{equation}

We must consider the possibility of anomalies arising from triangle diagrams in this model.  For a given triangle diagram with external gauge bosons $A^a_{\mu}, A^b_{\nu}, A^c_{\rho}$, the anomaly is proportional to 
\begin{equation}
\mbox{Tr}[(-1)^nT^a \{T^b,T^c\}]
\label{eq:genanom}
\end{equation}
where the $T^i$ are the generators corresponding to each of the gauge bosons, $n=0(1)$ for left-handed (right-handed) fermions, and the trace is over all fermions that can run in the loop.  A triangle diagram with exactly one $Z'$ boson does not give a triangle anomaly, as the trace over a single $SU(2)_F$ Pauli sigma matrix in (\ref{eq:genanom}) gives zero.  A diagram with three external $Z'$ bosons similarly vanishes. 

Thus, we need only consider those diagrams with two or zero $SU(2)_F$ gauge bosons.  For those diagrams with two $SU(2)_F$ gauge bosons, $Z_i'$ and $Z_j'$, those which contain a graviton or a SM $SU(3)$ or $SU(2)$ gauge boson are zero.  However, there is a constraint from the diagram with a hypercharge gauge boson; the anomaly for a diagram with an external $Z_i'$, $Z_j'$ and hypercharge gauge boson $B$ is proportional to
\begin{equation}
\mbox{Tr}[(-1)^n Y \{\tau^i,\tau^j\}]
\end{equation}
where $Y$ is hypercharge, and the $\tau^a$ are the $SU(2)_F$ Pauli sigma matrices.  Setting this to zero implies
\begin{equation}
\delta^{ij}\sum_{SU(2)_F} (-1)^n Y =0
\end{equation}
where the sum is over $SU(2)_F$ doublets.  With just $f$, $f'$, and the SM particle content, this relation is not satisfied, as the only nonzero term in the sum is $D_R$, with hypercharge $Y=-1/3$.  We can solve this problem by adding either another right-handed $SU(2)_F$ doublet with $Y=1$ or a left-handed $SU(2)_F$ doublet with $Y=-1$ which is a singlet under the other SM interactions.  (Here a factor of three arises since quarks carry color.)   However, we must also make sure we do not produce anomalies via diagrams which contain no $Z'$ gauge bosons, as this would spoil the anomaly cancellation of the SM.  We can achieve this by adding in two additional $SU(2)_F$ right-handed singlets, with hypercharge $Y=-1$.  We assume that all non-SM fields other than $f$ and $f'$ are sufficiently heavy to have escaped current experimental constraints. 

We now consider the possible contributions to $K^0-\bar{K}^0$ in this toy model.  At tree level, the operator ${\cal O}^R_{sdsd} =\bar{s}\gamma^{\mu}d\bar{s}\gamma_{\mu}d$ is not generated, as it does not obey the $SU(2)$ symmetry.  Additionally, loop diagrams which do not contain SM $W^{\pm}$ bosons or the DM particles $f, f'$ will not generate ${\cal O}^R_{sdsd}$.  Although one might expect this operator to possibly arise at one loop due to SM $W^{\pm}$ exchange, no contributions arise beyond those of the SM; as ${\cal O}^R_{sdsd}$ contains only down-type quarks, any diagram which contains only one $W^{\pm}$ boson must be such that the $W^{\pm}$ starts and ends on the same quark line.  However, as the $Z'$ does not couple to up-type quarks, no one-loop diagram with only one $W^{\pm}$ can be constructed which contributes to  ${\cal O}^R_{sdsd}$.  With two $W^{\pm}$ bosons, one recovers the usual SM $K^0-\bar{K}^0$ mixing contribution.  However, when we include the diagram shown in Fig. \ref{fig:kaonmix}, we obtain the results described in Sec. \ref{sec:flav}.  (Similar conclusions also apply to the $SU(2)_F$ doublet added in to cancel anomalies, as discussed above.)

\subsection{Vector couplings}

In this case, both the right-handed and left-handed quarks transform under the $SU(2)_F$ symmetry.  As the left-handed quarks come in $SU(2)$ weak doublets, we must include both the up-type and down-type quarks here.  Thus, we define, along with $D_R$, 
\begin{equation}
U_R=\left( \begin{array}{c}
u_{R1}\\
u_{R2}\end{array}\right)
\end{equation}
and
\begin{equation}
{\cal Q}_L=\left( \begin{array}{c}
Q_{L1}\\
Q_{L2}\end{array}\right)
\end{equation}
where the $Q_{Li}$ are the left-handed weak quark doublets.  Although we again take the interaction eigenstates to be quasi-aligned with the mass eigenstates, there is necessarily some mixing with the third generation, as the weak quark doublets are themselves not precisely aligned with the mass eigenstates.

We now consider anomalies for the case of vector interactions.  We again must consider the triangle diagram with two $SU(2)_F$ gauge bosons $Z_i'$ and $Z_j'$ and one hypercharge gauge boson.  In this case, the constraint is
\begin{equation}
\label{eq:anom}
\delta^{ij}\sum_{SU(2)_F} (-1)^n m Y
\end{equation}
where notation is as before, except that $m=1$ for $U_R$ and $D_R$, but $m=2$ for ${\cal Q}_L$, as  ${\cal Q}_L$ contains both up- and down-type quarks.  $Y(U_R)=2/3$, $Y(D_R)=-1/3$, and $Y({\cal Q}_L)=1/6$.  Thus, the sum in Eq. (\ref{eq:anom}) is zero without the addition of any particles beyond the SM.  Therefore, we need not consider diagrams containing only SM gauge bosons, as the anomalies cancel just as they do in the SM.

\begin{figure}[h]
\includegraphics[width=.3\textwidth, angle=0]{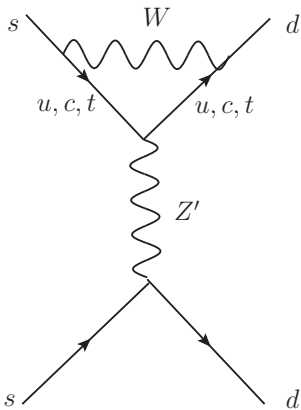}
\caption{$K^0-\bar{K}^0$ mixing contribution in the case of vector interactions.}
\label{fig:kaonmix2}
\end{figure}

We now consider possible contributions to $K^0-\bar{K}^0$ mixing.  Like in the right-handed case, ${\cal O}^V_{sdsd}$ is not generated at tree level.  However, it can be generated at one-loop level via the diagram shown in Fig. \ref{fig:kaonmix} or by a similar diagram with the dark sector particles replaced by quarks, or by the diagram in Fig. \ref{fig:kaonmix2}.  The diagrams containing quark loops are suppressed by two factors of SM quark mixing angles and by quark masses (with the largest contributions coming from $c$ quarks) and their contributions to $\Delta_{m_K}$ give limits on the new physics scale of $O(1 \mbox{ TeV})$.

We found in Section \ref{sec:flav} that $K^0-\bar{K}^0$ mixing implied a small mass difference $\delta$ or a small mixing angle between the mass and interaction eigenstates.  For vector interactions, the $SU(2)_F$ symmetry allows a bare mass term $m\bar{F}F$ without a coupling to $\varphi$.  Mass splitting and mixings can be accomplished, however, by coupling $\bar{F}F$ to $\varphi$ through higher-dimensional operators, such as $\bar{F}\varphi \varphi^{\dagger}F$.  As we would expect these operators to be suppressed by some mass scale, it may be reasonable to generate small mass splittings (even for large $m_f$) and/or small mixings in the dark sector.  

If we assume that such higher-dimensional operators give effective couplings of $\varphi$ to both our DM sector and the quark sector, we can have diagrams that allow $f'$ to decay, in addition to those studied in Sec. \ref{sec:direct}.  Here, we consider a diagram similar to that in Fig. \ref{fig:decay1} but with one of the $Z'$ bosons replaced with a scalar; we take the quark loop to contain a $c$ quark.  If we take the effective coupling $y_F$ of $\varphi$ to $\bar{F}F$ to be roughly $\sim\delta/v'$, and also assume that a similar relation holds for the quark mass splittings, such that $y_c\sim m_c/v'$, we estimate the decay width of the $f'$ to be
\begin{equation}
 \Gamma\sim\frac{1}{m_f}\left(\frac{\delta}{4\pi m_f}\right)\left(\frac{e^2m_c^2 y_F^2 y_c^2 m_f^2\delta^2}{(4\pi)^8\Lambda^4}\right)
\end{equation}
where the term in the first set of parentheses is a phase space factor and the term in the second set of parentheses comes from an order-of-magnitude estimate of the amplitude squared for this two-loop diagram.  For $\delta=1$ MeV and $v'\sim \Lambda\sim$ TeV, this gives $\Gamma \sim 10^{-50}$ GeV; for this mass splitting, the results of \cite{Yuksel:2007dr} exclude an $f'\rightarrow f\gamma$ for which $\Gamma\gtrsim 10^{-49}m_f$; thus, for rather small $m_f$, this diagram starts to probe an interesting region of the new physics scale.  We find no diagrams which give stronger constraints.  

\section{Signatures at LHC}
\label{sec:lhc}
We now consider the prospects for discovering flavored DM at LHC.  For the new physics scales which we have been considering, $O(\mbox{few TeV})$, the effective operator formalism which we have been using up until now is no longer applicable.  Here, we assume that flavor interactions are mediated by a heavy gauge boson, which we will label as $Z'$.  We note that the new physics scales which we have been considering include couplings, and that, for example, a new physics scale of $3$ TeV, which is allowed by $K-\bar{K}$ mixing and can still be accommodated by direct detection experiments, could easily correspond to a $1$-TeV $Z'$.  

For concreteness, we will consider the right-handed toy model described in Section \ref{sec:toy}.  This model contains three $Z'$ gauge bosons, all of which we will generically denote $Z'$, and all of which we will assume have a mass of $M_{Z'}=1$ TeV.  We will take the $f$ and $f'$ to have masses much less than $M_{Z'}$, and, for simplicity, we take the mixing angle between interaction and mass eigenstates to be zero.  All other fermions which are added into the model to cancel anomalies are assumed sufficiently massive that they are not accessible at LHC.  The $SU(2)$ coupling is taken to be the same as the SM $SU(2)$ coupling $g$.  These assumptions correspond to values of the effective operator coefficients
\begin{eqnarray}
\frac{|C^R_{ffdd}|}{\Lambda^2}=\frac{|C^R_{ffss}|}{\Lambda^2}=\frac{|C^R_{f'f'dd}|}{\Lambda^2}=\frac{|C^R_{f'f'ss}|}{\Lambda^2}&=&\frac{g^2}{4\mbox{ TeV}^2}\approx\frac{1}{(3\mbox{ TeV})^2}\nonumber\\
\frac{|C^R_{ff'sd}|}{\Lambda^2}=\frac{|C^R_{f'fds}|}{\Lambda^2}&=&\frac{g^2}{2\mbox{ TeV}^2}\approx\frac{1}{(2\mbox{ TeV})^2}
\end{eqnarray} 
with all other coefficients $C_{mnab}^R$ between the dark and SM sectors $0$. 

As an example of a possible search channel at LHC, we will consider the case where the $Z'$ is produced in conjunction with a jet, $pp\rightarrow Z'j$\footnote{Here, we consider all hard subprocesses of the forms $q_a g \rightarrow q_b Z'$, $\bar{q}_a g \rightarrow \bar{q}_b Z'$, and $q_a \bar{q}_b \rightarrow Z' g$ where $q_a$ and $q_b$ can be either $s$ or $d$.}, and decays invisibly to produce a monojet signature.  (For a discussion of monojet signatures at LHC, see \cite{Rizzo:2008fp}; for previous work on monojets with regards to DM, see \cite{Bai:2010hh}.)  An invisible decay of the $Z'$ can consist of $f\bar{f}$, but, in the case where $f'$ is adequately long-lived to leave the detector (which we assume here), we must also include the final states $f'\bar{f}$, $f\bar{f'}$, and $f'\bar{f'}$.  We use MadGraph/MadEvent \cite{Alwall:2007st} to calculate the cross-sections for these processes and for the SM backgrounds $pp\rightarrow Zj$ (with $Z\rightarrow \nu\bar{\nu}$) and $pp\rightarrow W^{\pm}j$ (where the $W^{\pm}$ decays leptonically and the charged lepton is lost down the beampipe, $\eta>2.5$).   We greatly reduce these backgrounds by placing a very tight cut on the transverse momentum $p_T$ of the jet.

\begin{figure}[h]
\includegraphics[width=.4\textwidth, angle=0]{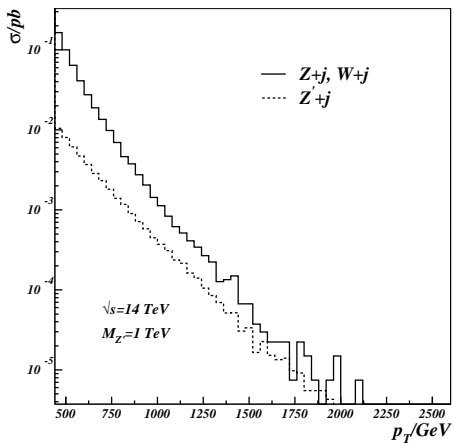}
\caption{Monojet $p_T$ distributions for a $1$-TeV $Z'$ signal and the SM background at $\sqrt{s}=14$ TeV.  A cut has been placed requiring $p_T>440$ GeV, which gives $S/B=10\%$.}
\label{fig:mc}
\end{figure}

The $p_T$ distributions for our $Z'$ signal and the SM background are shown in Fig. \ref{fig:mc} for a center-of-mass energy $\sqrt{s}=14$ TeV.  Although the two distributions are similar, the signal distribution falls off more slowly for large $p_T$.  Given the eventual expected data set for LHC (an integrated luminosity of $\sim 100 \mbox{ fb}^{-1}$), the discovery potential for such a $Z'$ would be expected to be limited by systematic errors, not by statistics.  Although a full study of systematics is beyond the scope of this paper, we note that a value of $S/B$ of $10\%$ can be obtained by requiring the monojet $p_T$ to be $>440$ GeV.  This cut has been applied in Fig. \ref{fig:mc}, giving signal and background cross-sections of $0.047$ pb and $0.47$ pb, respectively, and, for an integrated luminosity of $100 \mbox{ fb}^{-1}$, $S/\sqrt{B}\approx 22$.  $S/B=20\%$ can be achieved with a $p_T$ cut of $625$ GeV.  Although this is far from a complete analysis of this signature at LHC, these numbers indicate that this search channel merits further study.  

We briefly mention a few other possible signatures of this model at LHC.  First, we note that \cite{Petriello:2008pu} have studied the potential of discovering an invisibly-decaying $Z'$ produced in conjunction with an SM $Z$ which decays to leptons.  They conclude that a $5\sigma$ discovery of a $1$-TeV $Z'$ from a BSM $U(1)$ with gauge coupling of unity could be accomplished with $30 \mbox{ fb}^{-1}$ of data.  (See also \cite{Gershtein:2008bf} for a similar study of an invisible $Z'$ produced in conjunction with a $\gamma$.)  Additionally, visible decays of the $Z'$ could be considered.  In addition to $Z'\rightarrow jj$, we have, for example, $Z'\rightarrow f'\bar{f'}$; for $m_{f'}\gtrsim$ few GeV, and $m_f<<m_{f'}$, an $f'$ with a few hundred GeV of energy will decay in the detector, which can give, among other signals, $Z'\rightarrow \bar{f'} f'\rightarrow \bar{f} f jjjj$; depending on $m_{Z'}$, $m_{f'}$, and $m_f$, displaced vertices are also possible.  We do not consider these signals here, but mention that they could be studied in a more complete treatment.

\section{Conclusions}
\label{sec:conc}

We see that flavored DM is a rich subject.  Here, we have examined DM which interacts with quarks of the first two generations; we have then placed constraints on these interactions using low-energy measurements and direct detection, and considered the implications of the relic density for possible flavored DM models.  We also see that flavored DM has possible signatures at LHC and that it can give inelastic scattering (both up- and down-scattering) in direct detection experiments.  Throughout this analysis, we have strived to be as model-independent as possible.  

We collect some general results in Table (\ref{table:conc}).  Here, we show the approximate NP scale probed by each of these observables under the assumption of flavored DM.  We must emphasize that not all of these results will apply to all models, and, without a specific model, these numbers should not be compared to each other.  (For example, the results from $K^+$ decays are only applicable to very light DM.)  Additionally, these results can also be significantly altered in models with small mixing angles.  For these reasons, we give only order-of-magnitude estimates of the reach for each of these observables.  We do not include a number for $K-\bar{K}$ mixing as it depends very strongly on the mass splitting $\delta$.  Additionally, we have not done a complete study of the signatures of flavor $Z'$s at LHC, but we take a few TeV to be a reasonable estimate of the new physics reach for these scenarios.

\begin{table}[h]
\begin{tabular}{| c | c |}
\hline
Observable & Approximate NP scale reach\\
\hline
$K^+$ decays & $O(40-80\mbox{ TeV})$\\
\hline
$K-\bar{K}$ mixing & $\delta$-dependent\\
\hline
Relic density & few TeV\\
\hline
Direct detection (elastic) & $O(1-10\mbox{ TeV})$\\
\hline
LHC & few TeV\\
\hline
\end{tabular}
\caption{Order-of-magnitude estimates of the NP reach for various observables in flavored DM scenarios.  Note that, without a specific model, these numbers cannot be meaningfully compared with each other.  For more detailed information, see text.}
\label{table:conc}
\end{table}

Under the assumption that our particle $f$ comprises all of DM and couples to first-generation quarks, the constraints from direct detection are quite strong; in particular, in the case of vector couplings, only a very light $f$ would possibly be observable at LHC.  However,  considerations of the relic density indicate that additional interactions may be necessary; these additional interactions may introduce new signatures at LHC.  Thus, it may be fruitful to attempt to incorporate flavored DM into a more complete model.  If such a model contained multiple types of DM, the constraints on couplings to first-generation quarks from direct detection could also be loosened, opening up additional parameter space accessible at LHC. 

Finally, we note that we have confined ourselves to interactions involving $d$ and $s$ quarks (plus $u$ and $c$ quarks where necessary), and have limited ourselves to purely right-handed and purely vector couplings.  However, we would like to point out that the range of flavor interactions which could potentially be applied to DM is immense.  One could consider scalar interactions, interactions with leptons, and, perhaps most interestingly, interactions involving the third family of quarks.  This last option in particular could lead to interesting signatures in top physics at LHC.  We leave these ideas for future work.

\section{Acknowledgements}
The authors would like to thank H. Davoudiasl, H.-S. Lee, C. Lunardini, F. Paige, C. Sturm, and R. van de Water for helpful discussions and advice.  This work is supported under US DOE contracts No. DE-AC02-98CH10886 (BNL) and No. DE-FG02-91ER40684 (Northwestern).

\bibliographystyle{h-physrev}

\end{document}